\documentclass[twocolumn,showpacs,preprintnumbers,amsmath,amssymb]{revtex4}
\usepackage[dvips]{graphicx}
\usepackage{dcolumn}
\usepackage{bm}    

\begin{document}
\title{Coupled systems with hyperchaos and quasiperiodicity}
\author{Alexander~P.~Kuznetsov and Yuliya~V.~Sedova\footnote{Corresponding author.\\Tel.: +7 8452278685; fax: +7 8452272401.\\E-mail address: sedovayv@yandex.ru}}
\affiliation{Kotel'nikov's Institute of Radio-Engineering and Electronics of RAS, Saratov Branch,\\
Zelenaya 38, Saratov, 410019, Russian Federation\\
}

\date{\today}

\begin{abstract}
A model with hyperchaos is studied by means of Lyapunov two-parameter
analysis. The regions of chaos and hyperchaos, as well as autonomous
quasiperiodicity are identified. We discuss the picture of domains of
different regimes in the parameter plane of coupled systems, corresponding
to the cases of interaction of quasiperiodic and hyperchaotic subsystems.
\end{abstract}


\maketitle

A hyperchaotic attractor is characterized by the presence of several
positive Lyapunov exponents. Dynamics of systems with hyperchaos is more
complex and diverse than that of the system with simple chaos. Examples of
hyperchaos were pointed out yet by R\"{o}ssler and others researchers [1-6].
Recently some new suitable examples of models with hyperchaos were
discovered. A study of such problems as scenarios of the emergence of
hyperchaos, structure of the hyperchaotic attractor, a possibility of hidden
attractors et al. was undertaken in refs. [7-11]. It should be noted that
such systems are promising from the point of view of secure communication
because they can realize high-dimensional chaos. However, many questions in
the theory of hyperchaos remain open. For example, the problem of the
interaction of the hyperchaotic systems is important and promising. In this
report we discuss some aspects of this problem.

Let us consider a model suggested recently in paper [7]. It is a modified
form of well-known Lorenz system:
\begin{equation}
\label{eq1}
\begin{array}{l}
 \dot {x} = a(y - x) + eyz, \\
 \dot {y} = cx - dxz + y + u, \\
 \dot {z} = xy - bz, \\
 \dot {u} = - ky. \\
 \end{array}
\end{equation}
Here $x,y,z,u$ are the dynamical variables and $a,b,c,d,e,k$ are control
parameters.

The model (\ref{eq1}) has several fundamental properties and advantages [7]:
\begin{enumerate}
  \item it is dissipative at $\left( {a + b - 1} \right) > 0$;
  \item equations (\ref{eq1}) are symmetric with respect to z-axis: $\left( {x,y,z,u}
\right) \to \left( { - x, - y,z, - u} \right)$;
  \item it has only the trivial zero equilibrium.
\end{enumerate}

For fixed parameter values $a = 35,\;c = 25,\;d = 5,\;e = 35,\;k = 100$ Chen
et al. [7] undertook the one-parameter analysis of model (\ref{eq1}). In Fig.1 we
reproduced the spectrum of Lyapunov exponents and typical phase portraits at
varying parameter $b$. Note that, as usual, one exponent is always zero
$\Lambda _3 = 0$ and is not shown in the figure.

\begin{figure}[!ht]
\centerline{
\includegraphics[height=10.5cm, keepaspectratio]{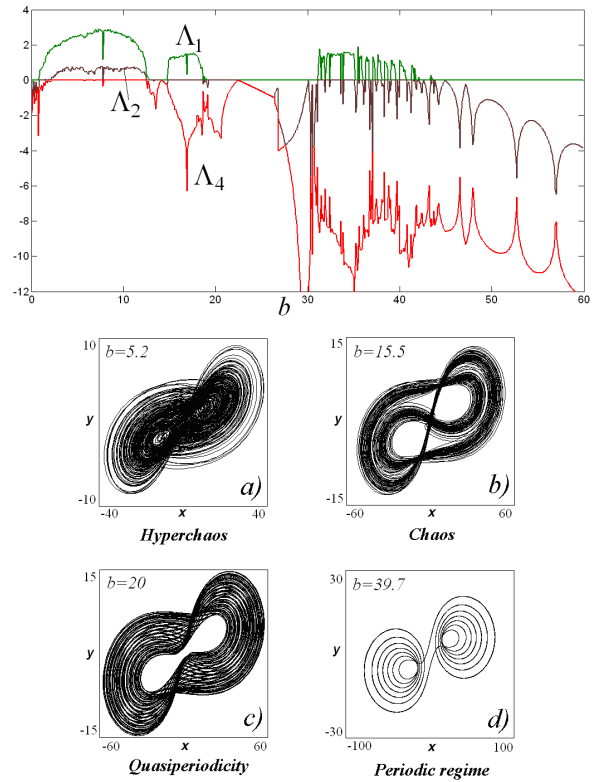}}
\caption{Plot of the Lyapunov exponents as a function of parameter $b$ for the
model (\ref{eq1}). (Exponent $\Lambda _3 = 0$ is not shown in the plot.)}
\end{figure}

The system (\ref{eq1}) exhibits hyperchaos in a wide range of parameter values. It
is interesting that along with such a regime, as well as periodic and
chaotic regimes, the model also demonstrates quasiperiodic dynamics. Thus,
model (\ref{eq1}) is one more example of an autonomous system with quasiperiodicity
(see. also [13,14]); this fact increases its importance.

One-parameter analysis does not give us full information about typical
regimes and their localization. That is why we turn to the two-parameter
analysis. Let us use charts of Lyapunov exponents [15-17]. We calculate the
values of the Lyapunov exponents at each point of the plane ($k,b)$ and
color this plane in accordance with the type of observed regime:
\begin{itemize}
  \item periodic regime $P$ (all exponents are negative);
  \item quasiperiodic regime $T_2 $ (one zero exponent);
  \item chaos $C$ (one positive exponent);
  \item hyperchaos \textit{HC} (two positive exponents).
\end{itemize}
\noindent(One trivial zero exponent is not taken into consideration.) The
corresponding chart of Lyapunov exponents is shown in Figure 2. The color
palette is presented to the right of the figure. It can be seen that the
area of hyperchaos entirely occupies extensive part of the plane of the
control parameters. The domain of existence of two-frequency tori is
visualized very well. We can distinguish the boundary of quasiperiodicity
region - the Neimark-Sacker bifurcation line with the Arnold tongues
emerging from it.

\begin{figure}[!ht]
\centerline{
\includegraphics[height=7cm, keepaspectratio]{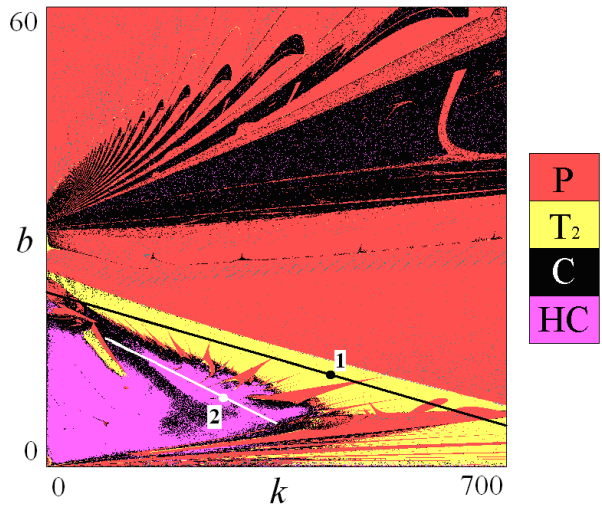}}
\caption{Chart of Lyapunov exponents of the system (\ref{eq1}). The color corresponds
to the type of regime. Two lines are shown for which the scan of parameter
plane is executed.}
\end{figure}

Let us study now dynamics of two coupled models (\ref{eq1}), the coupling terms are
represented by the differences of the corresponding variables:
\begin{equation}
\label{eq2}
\begin{array}{l}
 \dot {x}_1 = a(y_1 - x_1 ) + ey_1 z_1 + \mu \left( {x_2 - x_1 }
\right),\quad \quad \quad \quad \quad \\
 \dot {y}_1 = cx_1 - dx_1 z_1 + y_1 + u_1 + \mu \left( {y_2 - y_1 }
\right),\quad \quad \quad \,\,\,\;\, \\
 \dot {z}_1 = x_1 y_1 - b_1 z_1 + \mu \left( {z_2 - z_1 } \right),\quad
\quad \quad \quad \quad \quad \quad \;\;\;{\kern 1pt} \\
 \dot {u}_1 = - k_1 y_1 + \mu \left( {u_2 - u_1 } \right),\quad \quad \quad
\;\; \\
\\
 \dot {x}_2 = a(y_2 - x_2 ) + ey_2 z_2 + \mu \left( {x_1 - x_2 } \right), \\
 \dot {y}_2 = cx_2 - dx_2 z_2 + y_2 + u_2 + \mu \left( {y_1 - y_2 } \right),
\\
 \dot {z}_2 = x_2 y_2 - b_2 z_2 + \mu \left( {z_1 - z_2 } \right), \\
 {\kern 1pt} \dot {u}_2 = - k_2 y_2 + \mu \left( {u_1 - u_2 } \right). \\
 \end{array}
\end{equation}
Here $x_1 ,y_1 ,z_{1,} u_1 $ and $x_2 ,y_2 ,z_{2,} u_2 $ are dynamical
variables of the first and the second subsystems, respectively, and $\mu $
is the coupling parameter. We fixed a set of parameters $a = 35,\;c = 25,\;d
= 5,\;e = 35$ and change coefficients $k_1 ,k_2 ,b_1 ,b_2 $.

The presence of set of Arnold tongues of different periods in the autonomous
system allows us to associate with the model the so-called "frequency
parameter", corresponding to the motion along the Neimark-Sacker bifurcation
line. We choose two routes corresponding to varying of this effective
parameter value - one lies through the region of quasiperiodic regimes
characterized by built-in set of Arnold tongues (Route 1 in Figure 2) and
the second through the area of hyperchaos (Route 2).

These routes correspond to variation of the frequency parameter in the
second subsystem. In the first subsystem let us fix the values $k_1 ,\,b_1 $
so that we can observe either quasiperiodicity or hyperchaos. The values of
these fixed parameters are marked by points on lines in Figure 2.

\textit{The first scenario.} Let us fix parameters $k_1 = 430$, $b_1 = 12$ in the first subsystem and plot
a chart of Lyapunov exponents. Herewith along the y-axis we will vary
coupling parameter $\mu $ and along x-axis -- coefficient $k_2 $ measured
along the marked Route 1 in Figure 2. This case corresponds to interaction
of quasiperiodic oscillations. The resulting chart of Lyapunov exponents is
displayed in Figure 3. Now we have an additional possibility of the presence
of a three-frequency tori $T{ }_3$ (two zero exponents) and four-frequency
tori $T{ }_4$ (three zero exponents).

\begin{figure}[!ht]
\centerline{
\includegraphics[height=12cm, keepaspectratio]{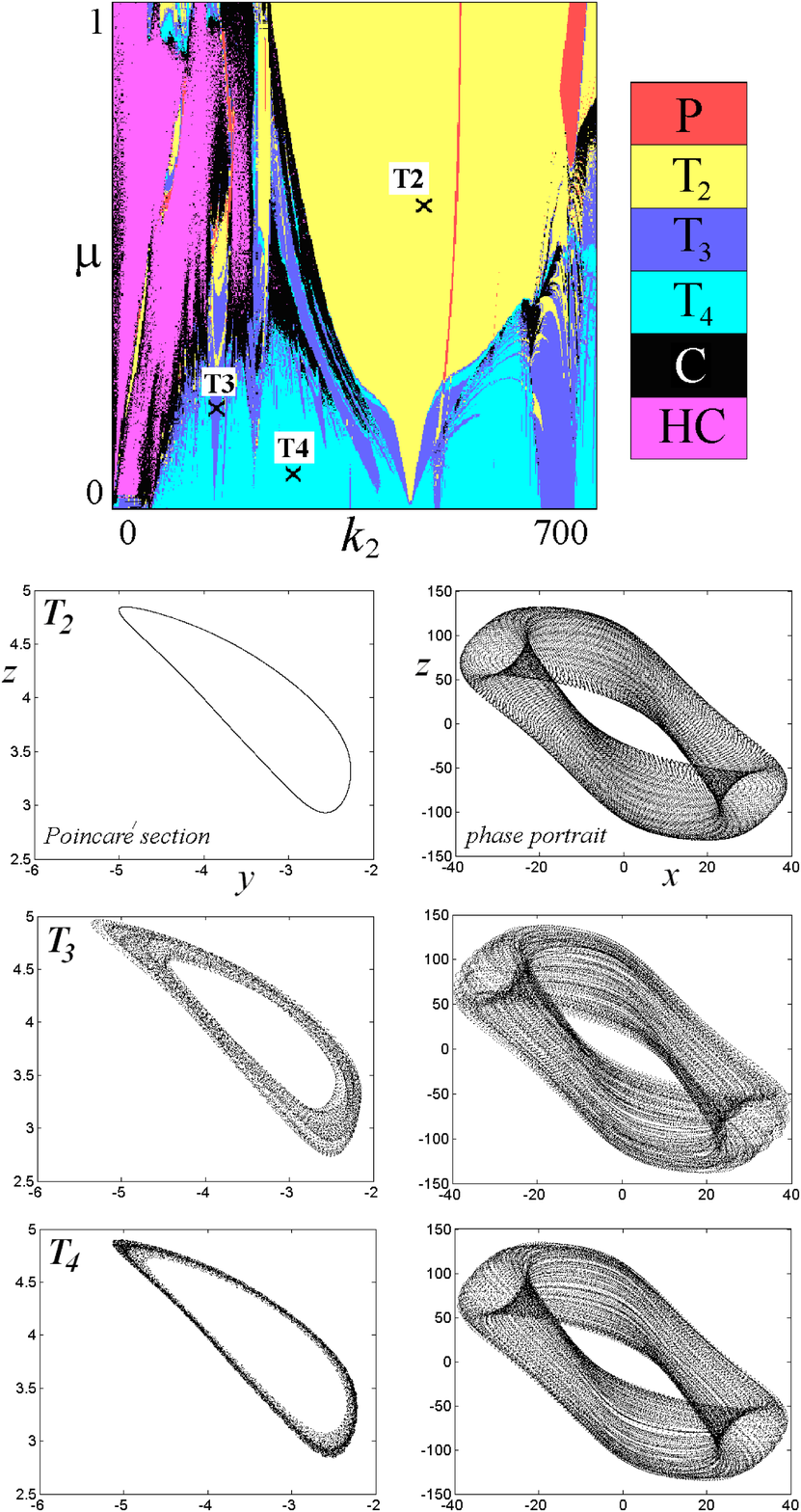}}
\caption{Chart of Lyapunov exponents for the system (\ref{eq2}), $a = 35$, $c = 25$, $d
= 5$, $e = 35$, $k_1 = 430$, $b_1 = 12$. Parameters $k_2$, $b_2 $ are varied along
the Route 1 in Figure 2. Poincar\'{e} sections and phase portraits in the
selected points on the Lyapunov chart are presented on the lower part of the
Figure.}
\end{figure}

At a low coupling parameter of the model (\ref{eq2}) it is naturally to expect an
emergence of four-frequency torus - this hypothesis is confirmed by the
numerical study. Also on the chart we can distinguish wide area of
two-frequency tori in form of peaked tongue touching the line of zero
coupling. The point of contact corresponds to the condition $k_2 = k_1
,\,\,b_2 = b_1 $ when the subsystems are identical and their "frequency
parameters" coincide. Also, there are narrow tongues of $T_3 $ torus
embedded in the region of $T_4 $ tori. In right part of Figure 3 we
demonstrate the phase portraits and Poincar\'{e} sections at points of
presence of $T_2$, $T_3$ and $T_4$ tori. A very narrow bands of periodic
regimes are visualized.

Note that inside the area of two-frequency tori it is possible to observe a
phase synchronization - some kind of \textit{synchronous quasiperiodicity}. Illustrations of such a regime are
given in Fig.4. We display the phase portraits of both partial systems of
model (\ref{eq2}), for which it is possible to determine the phase of the
oscillations (Figure 4a,b). Fig.4c represents the time-dependence of the
phase difference $\varphi _1 - \varphi _2 $; phase is calculated according
to the formula $\varphi _1 = \arctan \left( {{y_1 } \mathord{\left/
{\vphantom {{y_1 } {x_1 }}} \right. \kern-\nulldelimiterspace} {x_1 }}
\right),\varphi _2 = \arctan \left( {{y_2 } \mathord{\left/ {\vphantom {{y_2
} {x_2 }}} \right. \kern-\nulldelimiterspace} {x_2 }} \right)$. It can be
seen that the phase difference $\varphi _1 - \varphi _2 $ changes over time
in a limited range, which means the phase synchronization. Another proof of
the existence of phase synchronization is a specific form of the phase
diagram in Fig.4d. In Fig.4e we have shown time-dependence of the first and
second subsystems amplitudes $R_1 $ and $R_2 $.

\begin{figure}[!ht]
\centerline{
\includegraphics[height=7cm, keepaspectratio]{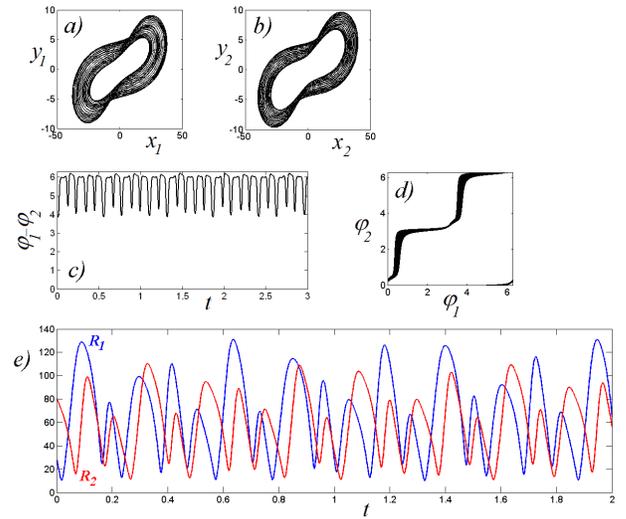}}
\caption{Illustrations of synchronous quasiperiodicity, $k_2 = 320$, $\mu = 0.7$.
a), b) -- phase portraits of both subsystems, c) the time-dependence of the
phase difference, d) phase diagram, e) the time-dependence of subsystem
amplitudes.}
\end{figure}

\textit{The second scenario.} Now in the second subsystem we will vary the parameters along the Route 2
in Figure 2. The first subsystem generates hyperchaotic oscillations; we
choose the selected point $k_1 = 430$, $b_1 = 12$ on this line. Thus, we are
going to discuss the case of interaction of two subsystems with hyperchaos.
Since every subsystem has two positive exponents in their own Lyapunov
spectrum, on Lyapunov chart it is logical to expect the area HC4 with four
positive Lyapunov exponents. This idea is confirmed by Fig.5a, that
demonstrates regimes with two, three and four positive Lyapunov exponents.
It is remarkable, that at large coupling not only the number of positive
Lyapunov exponents is reduced, but the regions of two-frequency
quasiperiodicity (although very narrow) arise. Thus, the interaction of the
hyperchaotic subsystems can lead to quasiperiodicity. This statement is
illustrated in Fig.5b, which presentes an enlarged fragment of the Lyapunov
chart. We can see the area of quasiperiodic oscillations and built-in
system of Arnold tongues.

\begin{figure}[!ht]
\centerline{
\includegraphics[height=8cm, keepaspectratio]{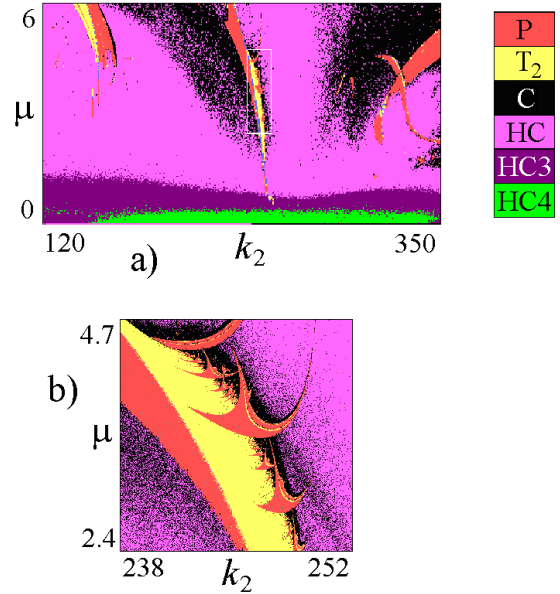}}
\caption{The chart of Lyapunov exponents of model (\ref{eq2}) and its enlarged
fragment.}
\end{figure}

Figure 6 is a plot of the largest four Lyapunov exponents along the line
$k_2 = const$ in Fig.5a. An interesting point is a noticeable change in the
irregularity of the plots in comparison with Fig.1. Thus, hyperchaos arising
from the interaction of subsystems is rather rough, i.e. weakly dependent on
parameters. This fact may be important for applications.

\begin{figure}[!ht]
\centerline{
\includegraphics[height=6cm, keepaspectratio]{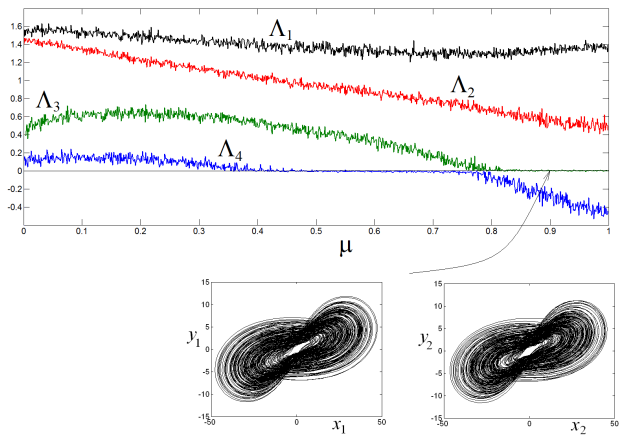}}
\caption{Plot of the largest Lyapunov exponents as a function of the coupling
parameter along the line $k_2 = 285$ in Fig.5. Phase portraits at $\mu =
0.9$ are presented.}
\end{figure}

Thus, in the present report we deal with the problems of the interaction of
the subsystems that can exhibit quasiperiodicity and hyperchaos in
autonomous mode. For coupled subsystems with autonomous quasiperiodicity we
detect the regime of synchronous quasiperiodicity. One of interesting
results is the emergence of quasiperiodic oscillations for the interacting
hyperchaotic systems.

\textit{This research was supported by the Grant of RF President program for leading Russian research school NSh-1726.2014.2 and Russian Foundation for Basic Research grant 14-02-00085.}

\end{document}